\documentclass[pra,amsmath,amssymb, twocolumn, showpacs]{revtex4}

\usepackage{dcolumn}

\usepackage{bm}

\usepackage[dvips]{graphicx}

\usepackage{epsfig}

\begin{document}

\title{Efimov states and their Fano resonances in a neutron-rich nucleus}

\author{I.\ M.\ Mazumdar$^{1,*}$, A.\  R.\  P. Rau$^{2}$, V.\ S.\ Bhasin$^{3}$}
\affiliation{$^{1}$Tata Institute of Fundamental Research, Mumbai 400 005, India, $^{2}$Department of Physics and Astronomy, Louisiana State University,
Baton Rouge, Louisiana 70803-4001, USA, $^{3}$ Department of Physics and Astrophysics, University of Delhi, Delhi 110 007, India}


\begin{abstract}

Asymmetric resonances in elastic n+$^{19}$C scattering are attributed to Efimov states of such neutron-rich nuclei, that is, three-body bound states of the n+n+$^{18}$C system when none of the pairs is bound or some of them only weakly bound. By fitting to the general resonance shape described by Fano, we extract resonance position, width, and the ``Fano profile index". While Efimov states have been discussed extensively in many areas of physics, there is only one very recent experimental observation in trimers of cesium atoms. The conjunction that we present of the Efimov and Fano phenomena may lead to experimental realization in nuclei.  

\end{abstract}


\maketitle


Over three decades ago, Efimov \cite{ref1} pointed out a spectacular effect, that a three-body system can support bound states under conditions when none of the three pairs constituting it are bound or one or two pairs are barely bound. Regardless of the nature of the pair-wise interactions, when any pair is on the threshold of binding (that is, large negative scattering length), an effective attractive, inverse-quadratic potential in the radial variable of the three-body system supports an infinite number of weakly bound states. More detailed studies followed \cite{ref2}, of the singularity structure near threshold and the motion of the states in the complex energy plane as a function of the strength of the pairwise interaction. Calculations in various atomic and molecular systems have also indicated the existence of Efimov states in helium and sodium trimers \cite{ref3}. Indeed, such ``Efimov physics" is now being seen as central to Bose-Einstein condensation and other ultracold phenomena in dilute atomic gases. And, very recently, since submission of this Letter, the first experimental observation of Efimov states has been reported in ultracold cesium trimers \cite{ref4}. Our study presented here of neutron rich nuclei has complementary aspects and holds the hope of observing this phenomenon in nuclear systems. For other discussions of halo states, including Efimov and other loosely bound configurations, see a review and references therein \cite{ref5}.

In a recent calculation of neutron rich nuclei, evidence was provided for a finite number of such Efimov states for weak pair binding, together with their appearance as resonances in an elastic scattering cross section as the Efimov states just become unbound \cite{ref6} . We now tie that result to yet another interesting but not fully familiar result, that a general resonance profile is asymmetric, as pointed out by Fano \cite{ref7} over four decades ago. While such asymmetric profiles have ben widely observed and studied in atoms and molecules, resonances in nuclear and particle physics generally show symmetric Lorentzian or Breit-Wigner shapes. We present asymmetric Fano resonances associated with Efimov states in a neutron-rich nucleus. Indeed, it is the combination of the features of the Efimov and Fano phenomena that holds promise for the possible observation of such resonances, with the asymmetry being used as a diagnostic for the Efimov effect. Indeed, our study as an Efimov bound state moves above threshold into the continuum is analogous to the recent observation in ultracold cesium which tracked a similar weakening of binding and disappearance as reflected in the loss rate of cesium atoms from an optical trap \cite{ref4}. With the advent of radioactive ion beams and other investigations of neutron-rich nuclei, the observation of the Efimov effect is likely and we sketch possible routes for experimental confirmation.
 
When studied as a function of energy, a scattering cross section exhibits a resonance when the energy traverses the position of a discrete state that is embedded in the energy continuum. Two alternative pathways to the final state, one directly into the continuum and the other through the embedded discrete state, interfere to give rise to the resonance. From general quantum-mechanical interference, such a profile may show both destructive and constructive features. Indeed, the most general profile embraces both as the energy traverses the resonance. Therefore, the general resonance profile may be expected to be asymmetric, with destructive and constructive interference on the two sides of the central resonance position. Fano \cite{ref7} presented the above picture in terms of interference between two competing amplitudes and developed the formalism for describing such resonances that has been widely adopted in atomic and molecular physics. Almost simultaneously and independently, Feshbach \cite{ref8} developed a formalism based on projection operators for describing so-called doorway states and other resonance phenomena in nuclear systems.

In atoms, resonances occur widely due to doubly and multiply excited states, such states often lying in the midst of a continuum built on ground or other lower-lying states \cite{ref9}. Such resonances have also been increasingly common in recent studies of exotic condensed matter systems, where they are sometimes associated with the Kondo phenomenon \cite{ref10}. Given its asymmetric profile, a resonance is described by three parameters, the energy position $E_r$, width $\Gamma$, and a so-called ``profile index" $q$ \cite{ref7, ref9}. Only when this last parameter becomes large does the profile reduce to the symmetric Breit-Wigner form that is more familiar to physicists as a resonance, especially in nuclear and particle physics \cite{ref11}. The reason for this reduction lies in $q$, which is the ratio of two quantities, the amplitude through the discrete state and the direct amplitude to the underlying continuum. In those instances, when the latter is small or negligible and the embedded discrete state is dominant, which is usually the case in nuclear and particle physics examples, the general profile reduces to Breit-Wigner, characterized by just two parameters, $E_r$ and $\Gamma$. 

\begin{figure}
\includegraphics[width=3in]{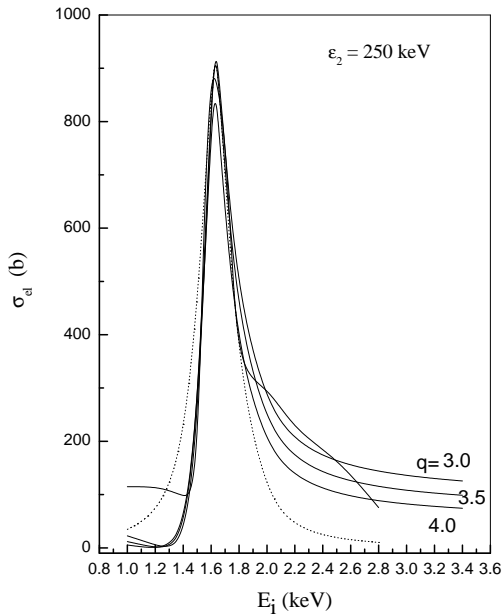}
\caption{
Elastic cross section for n-$^{19}$C scattering vs. centre-of-mass energy for a n-$^{18}$C binding energy of 250 keV. Calculated cross section (full curve) is fitted to the resonance formula in Eq.(1), the best fit obtained for parameters $E_r$ = 1.63 keV, $\Gamma$ = 0.25 keV, and $q$ = 4.0. The dotted line represents a Breit-Wigner fit to the calculated curve.}
\end{figure}

It is, therefore, interesting that a recent study \cite{ref6} of a nucleus showed a clearly asymmetric resonance profile. Fig. 1 from this study shows the elastic scattering cross-section for n+$^{19}$C at energies of a few keV. By tuning parameters for the two-body energy of n+$^{18}$C, just above the 220 keV when the first excited state of $^{20}$C becomes unbound, the above elastic cross-section exhibits a resonance as shown. Upon fitting these resonances to the Fano formula \cite{ref7,ref9,ref12},

\begin{equation}
\sigma = \sigma_0 [(q+\epsilon)^2/(1+\epsilon^2)],
\label{eqn1}
\end{equation}
where $\epsilon = (E-E_r)/(\Gamma /2)$ is a dimensionless, reduced energy measured from the central position in units of the width, and $\sigma_0$ the background cross-section far from the resonance, we get the parameters shown in Fig. 1.

This study \cite{ref6} of the dependence of the binding of the first excited state on underlying two-body parameters was done to exhibit the Efimov phenomenon \cite{ref1,ref2}. In such a study of two neutrons plus the $^{18}$C nucleus, we tune the two-body energy so as to make weakly bound states of $^{20}$C gradually get weaker in binding and then disappear to become quasi-bound states in the n+$^{19}$C continuum. For the first excited state, as shown in \cite{ref6}, this happens at about 220 keV. And, it is precisely at such energies above that value of 220 keV, when we have such an Efimov state embedded in the elastic scattering continuum of n+$^{19}$C, that we observe the resonance. 

\begin{figure}
\includegraphics[width=3in]{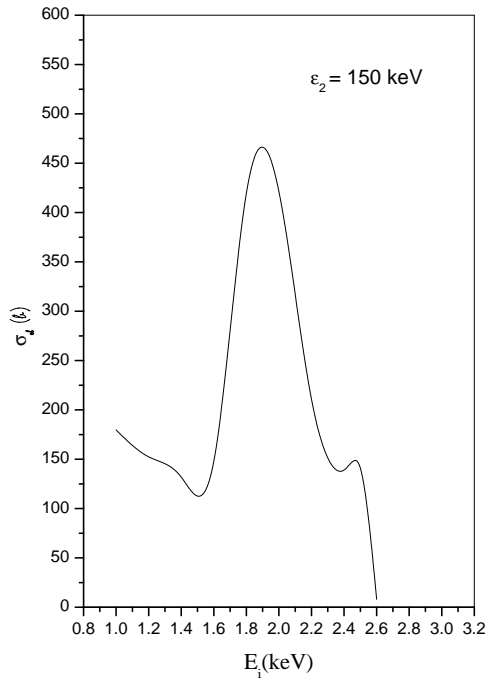}
\caption{Similar to Fig.\ 1 with lower binding energies of n-$^{18}$C, showing a resonance due to the second excited Efimov state. The 150 keV data fits Eq.(1) with approximately the same $q$ value as for the 250 keV curve in Fig.\ 1.}
\end{figure}

In turn, this suggests a similar effect for the second excited Efimov state which was seen \cite{ref6} to disappear above the n+$^{18}$C threshold at about 140 keV. Indeed, we verified this expectation as shown in Fig. 2. A Fano resonance appears at energies above 140 keV, pointing to the generality of the phenomenon. The very weakness of the binding and large spatial spread of these Efimov states lead to strong overlap with continuum states. This makes for comparable amplitudes of the two pathways and thus accounts for the very asymmetric profile. We note that the two resonances in Figs.\ 1 and 2 have approximately teh same $q$ value, displaying their origin as members of the same family of Efimov states. Loosely bound Efimov configurations in other neutron-rich nuclei such as $^{19}$B and $^{22}$C \cite{ref13} may also show similar Fano resonances. Three Efimov states were observed in $^{19}$B for a scattering length of -179 fm. The clearest demonstration of the Efimov phenomenon would, of course, be to observe a series of such states $n$ in the limit of very large scattering length, with energy positions relative to threshold and widths scaling with a characteristic exponential dependence on $n$ \cite{ref14}. However, that very exponential dependence and realistic values of scattering lengths mean that very few states are usually seen in practice and also, that there are corrections to that exponential dependence.

A more detailed study of such universal scaling with $n$ will have to await a better knowledge of n+$^{18}$C binding. Even without a rigorous examination of realistic nuclear potentials, we note that plausible arguments have been given for universality in the n+n+$^{18}$C system \cite{ref14}. Further aspects of universality, studies of limit cycles, and corrections because of finite scattering length will have to await much better understanding of these neutron-rich systems.  For now, we have relied on other ways such as the tuning of the two-body energy to pass from bound to continuum character and the observation of an asymmetry in the profile as our diagnostics. Our obtaining the same value of $q$ for the two resonances is in conformity with their constituting a sequence, and lends further support to our Efimov interpretation. Interestingly, the recent experiment with ultracold atoms \cite{ref4} also saw only a couple of Efimov states and again, just as in our study, through their disappearance as the scattering length was tuned through threshold (in that case, by changing the magnetic field), without exhibiting the universal exponential scaling with $n$.         

\begin{figure}
\includegraphics[width=3in]{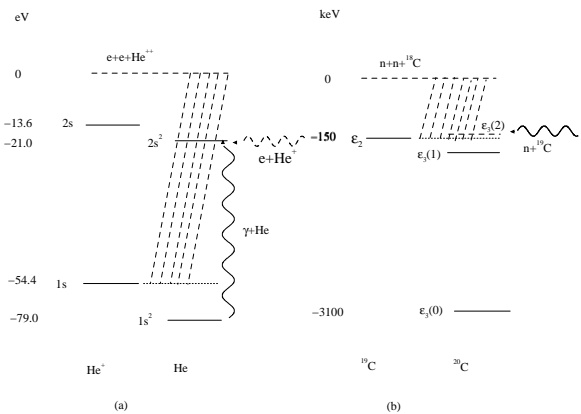}
\caption{
Comparison between He and $^{20}C$ as three-body systems in atoms and nuclei. (a) Ground and first excited state (there are infinitely many) of He$^+$ and the lowest bound states of He (again there are infinitely many) attached to them are shown, along with their energies below the fully dissociated limit of two electrons and the helium nucleus. Cross-hatched region is the ionization continuum of He built on the ground state of He$^+$, that is, the two-electron states $1sEs$, $E$ being the energy of the continuum electron. The $2s^2$ state lies embedded in this continuum, that is, is degenerate with $1sEs$ and, therefore, mixes with it to give the quasi-bound resonance state. The resonance can be accessed either by (e + He$^+$) scattering or by photoexcitation from the ground state as shown (because of dipole selection rules, a single photon will reach similar states of $2s2p \,^{1}P^o$ symmetry whereas two-photon absorption would be necessary to reach the $2s^2$). (b) Similar schematic for the nuclear system. Now $^{19}$C has only one bound state, its ground state $\epsilon_2$. For an appropriate value of $\epsilon_2$, one bound Efimov state $\epsilon_3 (1)$ and a second which just fails to be bound, $\epsilon_3 (2)$, are shown, together with the ground state of $^{20}$C. The state $\epsilon_3 (2)$, lying embedded in the n+$^{19}$C continuum, manifests as a resonance in low energy elastic scattering of n+$^{19}$C. For somewhat more negative $\epsilon_2$, the first Efimov state $\epsilon_3 (1)$ also fails to be bound and will appear as a resonance in n+$^{19}$C scattering as in Fig.\ 1.} 
\end{figure}

It is also instructive to contrast these nuclear three-body systems with the somewhat analogous doubly excited states of the helium atom. Fig.\ 3 provides such a schematic, singling out the He $2s^2 \,^{1}S$ state as an example \cite{ref15}. This is a doubly-excited state, with both electrons excited out of the ground $n=1$ quantum number. It lies embedded, as shown, in the $1s Es \, ^{1}S$ continuum ($E$ is the energy of the continuum electron) and mixes with it to give the resonance state. Note that it lies, approximately, 60 eV above the ground state of helium, which is enormous in the scale of atomic energies, the first ionization potential of He being 24.6 eV. The resonance would be seen in (e+He$^{+}$) scattering at kinetic energies of about 33 eV. This is in exact analogy with our nuclear example shown alongside in Fig.\ 3.

Differences between the atomic and nuclear systems are noteworthy. Being an (attractive) Coulomb system, He$^{+}$ has an excited bound state $2s$ lying 40.8 eV above He$^{+} 1s$, and the $2s^2$ state is bound relative to it. (Similarly, there are an infinity of other doubly excited states below an infinity of excited states of He$^{+}$). With short range forces between the neutron and $^{18}$C, there is no counterpart excited state of $^{19}$C and bound states below it to serve as embedded states in the n+$^{19}$C continuum. It is, therefore, the Efimov effect which provides such a candidate for the subsequent Fano resonance we describe. Unlike the atomic case, with the long range Coulomb force readily providing an infinite number of doubly-excited states and corresponding Fano resonances, it is the very occurrence of the Efimov state arising from weak, attractive, pair-wise forces that makes possible such a resonance in the nuclear system. There is  a pleasing interplay between the two phenomena, the loose binding of the Efimov effect being responsible for the substantial overlap between two pathways that gives rise to an asymmetric Fano profile and that asymmetry being the diagnostic for the Efimov phenomenon.

Finally, the analogy provided in Fig.\ 3 suggests a possible alternative pathway for observing the Fano resonance instead of elastic scattering of neutrons, namely, photoexcitation. Just as in the atomic counterpart, where photoabsorption from the ground state with photons of energy 60 eV (approximately) also accesses doubly-excited states (although not of $^{1}S^e$ but of $^{1}P^o$ symmetry because of dipole selection rules), we can expect that absorption of gamma rays from lower states such as the ground state of $^{20}$C will show Fano resonances from the Efimov states in Fig.\ 3. A further analogy with the observation of doubly-excited states through collisions of He with heavy charged particles such as positive ions provides yet another route for observing the Efimov-Fano resonances in nuclei, namely, fragmentation of $^{20}$C on a heavy target. With the advent of new machines for generating and studying very neutron-rich nuclei, we can look forward to the observation in nuclei of the effect predicted by Efimov many years ago.  

ARPR's work has been supported by the U.S. National Science Foundation Grant 0243473 and by the Roy P. Daniels Professorship at LSU. He also thanks the Tata Institute of Fundamental Research for its hospitality during the course of this work.

\end{document}